\documentclass{article}
\usepackage{ijcai23}

\usepackage{times}
\usepackage{soul}
\usepackage{url}
\usepackage[hidelinks]{hyperref}
\usepackage[utf8]{inputenc}
\usepackage[small]{caption}
\usepackage{graphicx}
\usepackage{amsmath}
\usepackage{amsthm}
\usepackage{booktabs}
\usepackage{algorithm}
\usepackage{algorithmic}
\usepackage[switch]{lineno}
\usepackage{multirow}

\usepackage{xcolor}

\title{Perturbing a Neural Network to Infer Effective Connectivity: Evidence from Synthetic EEG Data}

\author{
Peizhen Yang$^1$\footnote{Co-first authors: P.Y, and X.S.}
\and
Xinke Shen$^{1,*}$
\and
Zongsheng Li$^{1,2}$
\and
Zixiang Luo$^{1}$
\and
Kexin Lou$^{1,3}$
\and
\\
Quanying Liu$^{1}$\footnote{Corresponding author: Q.L., liuqy@sustech.edu.cn}
\affiliations
$^1$Department of Biomedical Engineering, Southern University of Science and Technology\\
$^2$Department of Computer Science, University of Macau\\
$^3$School of Information Technology and Electrical Engineering, University of Queensland\\
}

\begin{document}

\maketitle

\begin{abstract}
   Identifying causal relationships among distinct brain areas, known as effective connectivity, holds key insights into the brain's information processing and cognitive functions. Electroencephalogram (EEG) signals exhibit intricate dynamics and inter-areal interactions within the brain. However, methods for characterizing nonlinear causal interactions among multiple brain regions remain relatively underdeveloped. In this study, we proposed a data-driven framework to infer effective connectivity by perturbing the trained neural networks. Specifically, we trained neural networks (\textit{i.e.}, CNN, vanilla RNN, GRU, LSTM, and Transformer) to predict future EEG signals according to historical data and perturbed the networks' input to obtain effective connectivity (EC) between the perturbed EEG channel and the rest of the channels. The EC reflects the causal impact of perturbing one node on others. The performance was tested on the synthetic EEG generated by a biological-plausible Jansen-Rit model. CNN and Transformer obtained the best performance on both 3-channel and 90-channel synthetic EEG data, outperforming the classical Granger causality method. Our work demonstrated the potential of perturbing an artificial neural network, learned to predict future system dynamics, to uncover the underlying causal structure.
   
\end{abstract}

\section{Introduction}

The brain is a profoundly intricate, interwoven network characterized by causal influences among its various regions~\cite{sporns2004organization,van2010exploring,liu2017detecting}. From brain recordings like functional magnetic resonance imaging (fMRI) or electroencephalogram (EEG), one can easily compute the correlation of signal dynamics across different regions, which is typically referred to as functional connectivity (FC)~\cite{park2013structural,van2010exploring,samogin2019shared}. However, FC cannot characterize the \textit{directionality} and the \textit{sign} of connectivity~\cite{buckner2013opportunities}. Instead, effective connectivity (EC) can reflect the underlying directional causal influence from one brain region to another with the strength, directionality, and sign~\cite{luo2022effective,kim2023wholebrain}. It characterizes the information flow across brain regions and is critical for understanding how information is integrated or segregated during the cognitive process. Therefore, developing methods to reliably estimate EC from the recorded neural data stands as a significant endeavor in the field of neuroscience~\cite{Reid2019Advancing}.

\begin{figure}[t]
\centering
\includegraphics[width=1\linewidth]{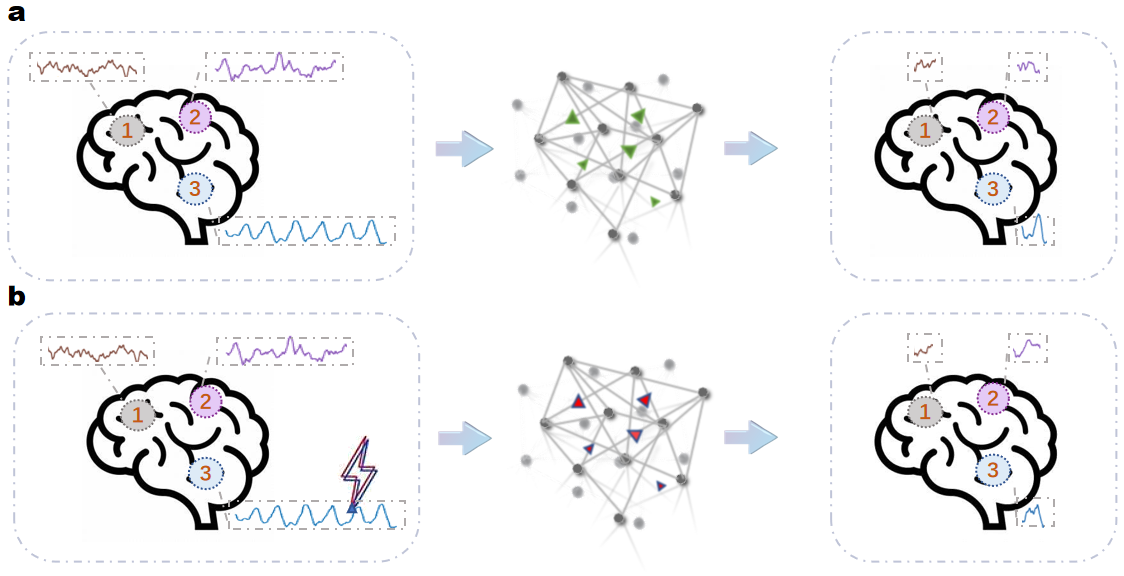}
\caption{\textbf{Framework of perturbation-based EC inference on EEG data.} \textbf{a}, Training artificial neural networks (ANNs) for EEG signal prediction. ANNs are trained to predict EEG signals of the following time steps from multiple previous steps. \textbf{b}, Perturbing the trained ANNs to infer EC. The trained ANNs are used as surrogate brains. EC is estimated by sequentially perturbing each region of the surrogate brain and measuring the stimulation-induced responses.}
\label{fig:framework}
\end{figure}

A number of computational methods have been proposed to investigate effective connectivity among multiple brain regions based on time series of neural signals, including Granger causality(GC) ~\cite{granger1969investigating} and dynamic causal modeling (DCM)~\cite{friston2003dynamic}. However, these methods are limited in their capacity to accurately encapsulate complex, non-linear interactions. 
For example, Granger causality operates on the premise that if the past activities of brain region A can predict the activities of another region B, there should be a causal interaction from region A to B. Granger causality typically relies on the assumptions of linear dependency and is sensitive to the effect of noise~\cite{friston2014granger}. On the other hand, DCM employs a Bayesian framework to identify the nonlinear input–state–output systems from observed data~\cite{friston2003dynamic,penny2004comparing}. Despite its popularity in neuroscience, DCM is limited by the pre-designed system dynamic model and its computational demand with an increasing number of nodes~\cite{lohmann2012critical,daunizeau2011dynamic}.

Rather than identified with computational methods, effective connectivity can be directly mapped with in-vivo experiments by applying electrical or optical stimulations to a specific brain region and then evaluating the resultant impact on other regions~\cite{kim2023wholebrain}. Although such perturbation-based experimental approaches are straightforward, they are typically unfeasible in human subjects due to ethical reasons and technical limitations. To address this challenge, we developed a data-driven framework that leverages a surrogate brain model to capture brain dynamics. Within this framework, we apply in-silico perturbations to various brain regions and observe the subsequent influence on other regions. This conceptually simple approach allows us to investigate effective connectivity by virtual perturbations. 

Here, we employed artificial neural networks (ANNs) to characterize the nonlinear interactions among brain regions. By training a neural network to predict the future dynamics of neural signals, we create a surrogate brain that could be perturbed to uncover causal interactions (Fig.~\ref{fig:framework}. This approach enables us to leverage state-of-the-art ANN architectures in time series forecasting and comprehensively investigate the optimal characterization of intricate nonlinear relationships among brain regions.

In our experiment, we used synthetic EEG data generated by a biological-plausible Jansen-Rit (J-R) mmodel~\cite{coronel2021cholinergic}, which can capture the fast-changing nonlinear dynamics of EEG. As the ground-truth effective connectivity is unknown in real EEG data, we developed a testbed for the framework using synthetic data. The ground-truth effective connectivity can be obtained from the synthetic data by perturbing the hidden variables during data generation. We generated two datasets: 1) A simple J-R synthesized dataset with 3 regions and pre-defined connections as a proof-of-concept; 2) A J-R synthesized dataset with 90 regions, in which the connections were real structural connectivity measured from diffusion tensor imaging (DTI).

The contributions of this paper are two-fold:

\begin{itemize}
    \item We presented a testbed for verifying the data-driven EC inference framework on fast-changing synthetic EEG data with known real EC. Various neural network models were tested in this testbed.
    \item We validated the effectiveness of the neural perturbational inference framework in comparison to classical EC estimation methods. The results underline the importance of selecting a proper model to serve as a surrogate brain.
\end{itemize}

\section{Problem statement}
The primary aim of this work is to estimate the causal influence of one brain region on others. To achieve this end, we implement virtual perturbation on the trained neural networks that can predict the dynamics of neural signals. The prediction model can be represented as 
\begin{equation}
\hat{\mathbf{x}}_{t+1: t+T'}=f({\mathbf{x}}_{t-T: t}, \theta), 
\end{equation}
which means predicting neural activities $\hat{\mathbf{x}}_{t+1: t+T'}$ with a nonlinear model $f$ based on previous activities $\mathbf{x}_{t-T: t}$. After the model is trained, we perturb one region at a time and see the changes in the predicted signals of other regions, which can be formulated as
\begin{equation}
\begin{aligned}
\delta_{A \rightarrow B}(t+t')= \mathbf{E}{[\left(B_{t+t'} \mid A_t+\Delta\right)-\left(B_{t+t'} \mid A_t\right)]}, \\
t'=1,2,...,T'
\end{aligned}
\end{equation}
\\
Here, we add perturbation $\Delta$ on region $A$ at time $t$ and see the expected changes in the region $B$ at time $t+1$ to $t+T'$. We perturb every region in a loop and see the causal influence between any two region pairs in this way.

\section{Related Work}

\subsection{Classical EC estimation methods}

The classical computational methods for estimating effective connectivity from neural data can be broadly categorized based on two aspects: i) linearity/nonlinearity and ii) bivariate/multivariate analysis. For instance, Granger causality, as a linear bivariate method, usually focuses on the linear interactions between two brain regions~\cite{friston2014granger}. Transfer entropy, as a nonlinear bivariate method, quantifies the directed transfer of information between two regions~\cite{yang2012new}. On the other hand, multivariate models take into account the interactions among multiple regions concurrently. These multivariate models have the advantage of mitigating spurious connections that may arise in bivariate models. For example, the partially directed coherence~\cite{baccala2001partial} and directed transfer function~\cite{wilke2008estimation} derive causal interactions from the Fourier transform of multivariate autoregressive parameters. The capability of these methods to simultaneously identify nonlinear multivariate interactions are underdeveloped. 
Recently, deep neural networks have shown their great expressive power for multivariate time-series prediction~\cite{wang2019multiple,bianchi2020reservoir,liang2022online}. However, whether their expressive power can transfer to accurately capture the complex and nonlinear interactions in the multivariate data requires further investigation.

\subsection{Neural perturbational inference}

The data-driven framework of Neural Perturbational Inference (NPI) was proposed by Luo et al.~\cite{luo2022effective}. NPI uses an artificial neural network (ANN) that learns neural dynamics as a surrogate brain. Perturbing the surrogate brain (\textit{i.e.}, the trained ANN), region by region, and simultaneously observing the evoked neural response at all unperturbed regions provides the whole-brain effective connectivity. The ANN in NPI is instantiated with a four-layer perceptron and is trained using a one-step-ahead prediction error, where the next state of fMRI is predicted given the current state. After ANN is trained, each region of ANN is sequentially perturbed, realized as a small increase or decrease of neural signal in the perturbed region. The EC is computed by the difference between the one-step neural responses with and without perturbation. 

The NPI framework has demonstrated its ability to infer EC from fMRI signals~\cite{luo2022effective}. Owing to the long timescale and slow dynamics in fMRI signals, the current state contains most of the useful information to effectively forecast the next state~\cite{Nozari2021brain}. Therefore, the ANN in NPI is simply realized with a multi-layer perceptron trained using one-step-ahead prediction error~\cite{luo2022effective}. However, this one-step prediction approach may fall short for capturing EEG dynamics. As the nature of EEG dynamics is highly nonlinear and complex, with rich information for predicting the next EEG signal is contained in many previous steps. Therefore, NPI developed for fMRI data cannot be directly applied to EEG. Here, we extended the original NPI framework with two factors: i) the ANN in the NPI framework to predict EEG dynamics is replaced with the state-of-the-art time-series prediction models (\textit{e.g.}, RNN, GRU, LSTM, CNN, and Transformer), ii) the EC is estimated with the multi-step response after perturbation, rather than one-step transient response.

\section{Methods}

\subsection{The framework}

The framework of our method is shown in Fig. 1. To learn the system dynamics from EEG signals, we trained a time series forecasting model to predict the subsequent signals using previous $n$ steps of EEG data (Fig. 1(a)). The virtual perturbation was applied to a region at the $76^{th}$ step. (Fig. 2(b) left), and the responses across all brain regions at future $77$ to $99$ steps were predicted by the trained models (Fig.2(b) right). The estimated EC was calculated as the difference between the expected signals with and without disturbance. To obtain the whole-brain causal connection between any two regions, we individually make an impulse perturbation (unit=0.1) into each region. 

\subsection{The time series forecasting models}
Following the instruction of the NPI framework, the EC inference largely relies on the time series forecasting model. In this work, we realized five artificial neural network models (CNN, vanilla RNN, LSTM, GRU and transform) as the EEG series forecasting model. 
The models are detailed in the Appendix.

\subsection{Synthetic EEG data}

We used the biologically plausible Jansen-Rit model, as shown in Eq.\eqref{eq:JR}, to generate EEG data. The Jansen-Rit model is a mathematical model used to simulate the macroscopic electrical behavior observed in EEG signals. The simulated data mimic the real EEG data in nonlinear dynamics and complex inter-regional interactions. For each brain region, it assumes three populations of neurons: pyramidal neurons, excitatory interneurons, and inhibitory interneurons. Pyramidal neurons have projections to the other two populations. Excitatory and inhibitory interneurons project back to pyramidal neurons. The pyramidal neurons also have long-range excitatory projections to other brain regions. The dynamics of each region are represented as follows:

\begin{small}
\begin{equation} \label{eq:JR}
\begin{aligned}
\dot{x}_{0, i}(t)&=y_{0, i}(t) \\
\dot{y}_{0, i}(t)&=A a\left[S\left(C_2 x_{1, i}(t)-C_4 x_{2, i}(t)+C \alpha z_i(t), r_0\right)\right]\\
& -2 a y_{0, i}(t)-a^2 x_{0, i}(t) \\
\dot{x}_{1, i}(t)& =y_{1, i}(t) \\
\dot{y}_{1, i}(t)& =A a\left[p(t)+S\left(C_1 x_{0, i}(t)-C \beta x_{2, i}, r_1\right)\right]\\
& -2 a y_{1, i}(t)-a^2 x_{1, i}(t) \\
\dot{x}_{2, i}(t)& =y_{2, i}(t) \\
\dot{y}_{2, i}(t)& =B b\left[S\left(C_3 x_{0, i}(t), r_2\right)\right]-2 b y_{2, i}(t)-b^2 x_{2, i}(t) \\
\dot{x}_{3, i}(t)& =y_{3, i}(t) \\
\dot{y}_{3, i}(t)& =A \bar{a}\left[S\left(C_2 x_{1, i}(t)-C_4 x_{2, i}(t)+C \alpha z_i(t), r_0\right)\right]\\
& -2 \bar{a} y_{3, i}(t)-\bar{a}_i^2 x_{3, i}(t)
\end{aligned}
\end{equation}

\end{small}
where $x_0$, $x_1$ and $x_2$ represent the output of the pyramidal neurons, excitatory interneurons, and inhibitory interneurons, respectively. $x_3$ represents the long-range output of the pyramidal neurons to other regions. $S$ is a sigmoid function:

\begin{equation}
S(v, r)=\frac{\zeta_{\max }}{1+e^{r(\theta-v)}}
\end{equation}
$z_i$ is the overall input from other regions to region $i$:

\begin{equation}
z_i(t)=\sum_{j=1, j \neq i}^n \widetilde{M}_{i j} x_{3, j}(t)
\end{equation}
where $\widetilde{M}_{i j}$ is the normalized structural connectivity matrix:

\begin{equation}
\tilde{M}_{i j}=\frac{M_{i j}}{\sum_{j=1, j \neq i}^n M_{i j}}
\end{equation}
$M_{i j}$ represents the underlying structural connectivity from region $j$ to region $i$. The EEG-like signal is calculated as:
\begin{equation}
v_i(t)=C_2 x_{1, i}(t)-C_4 x_{2, i}(t)+C \alpha z_i(t)
\end{equation}
which represents the postsynaptic potentials of pyramidal neurons in region $i$.

We generated two versions of synthetic data: 1) A toy example with 3 nodes. The structural connectivity among the 3 nodes was set manually, with node 0 exerting directed connections to node 1 and node 2 (Fig. 2a). 2) A whole-brain model with 90 nodes. Connectivity among the nodes was determined by real structural connectivity measured from DTI. The brain was parcellated into 90 regions with Anatomical Automatic Labeling (AAL) atlas. The structural connectivity was calculated from the average of 245 subjects in the Human Connectome Project (https://www.humanconnectome.org/study/hcp-young-adult/document/1200-subjects-data-release) (Fig. 3(a,b)).

\paragraph{Hyperparameter settings of Jasen-Rit model.} In this study, the excitatory gain $\alpha$ and the inhibitory gain $\beta$ were set as 0.71 and 0.4, respectively, to generate reasonable power spectra and functional connectivity properties in synthetic data. All other hyperparameters were set the same as in ~\cite{coronel2021cholinergic}. 
The neural dynamic described in Eq.(3) was transformed into a discretized formula by the forward-Euler method and evolved with a time step of 0.001 seconds. Then the generated EEG signals were downsampled to 100 Hz, with a time interval of 0.01 seconds between two consecutive time steps.

\begin{figure}[h]
\centering
\includegraphics[width=1\linewidth]{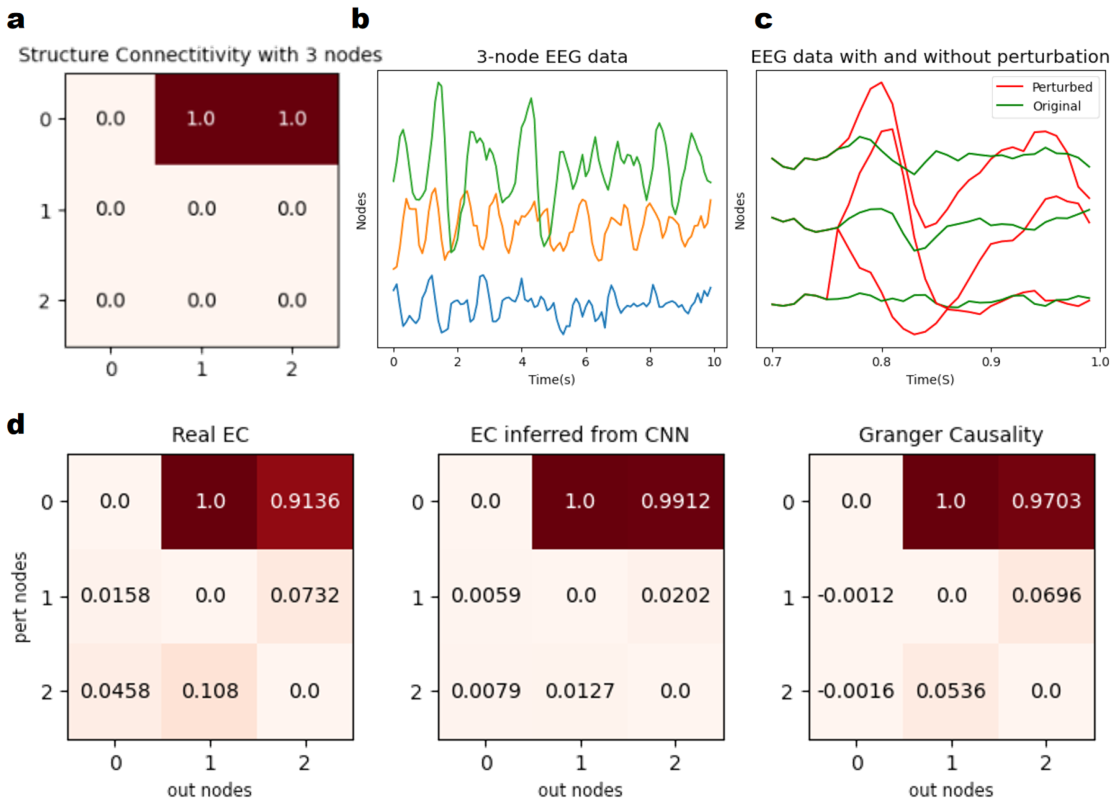}
\caption{\textbf{Data and results visualization of the 3-channel synthetic EEG}. \textbf{a}, The setting for 3-channel structural connectivity matrix. \textbf{b}, An example of 3-channel EEG data. \textbf{c}, The time course of the original EEG data (the green lines) and the perturbed EEG data (the red lines). \textbf{d}, the comparison among the real EC (left), the EC inferred by perturbing a CNN model (middle), and EC inferred by Granger causality (right). All the EC matrices were re-scaled to the range 0-1 for visualization.}
\label{fig: three node result}
\end{figure}

\subsection{Implementation details of ANN models}

We implemented five ANN models, including temporal CNN, vanilla RNN, LSTM, GRU, and Transformer models. Each ANN model consists of two hidden layers with 8, 32, 128, and 512 units, respectively, thus we can examine the impact of the model complexity on the performance of data prediction and EC inference. A linear readout layer was employed to predict future EEG dynamics. We applied Adam optimizer and ReduceLROnPlateau scheduler while training. The initial learning rate was $1e^{-4}$ and the batch size was 30. The number of training epochs was determined according to the minimum validation loss of each ANN model. 

\paragraph{Training and testing datasets.} 
For training and validating the forecasting model, an EEG signal with 900,000 time points was generated. The first 70\% of the generated time series was used in training and the remaining 30\% was used in validation. Each training or validation sample contains signals of 100 time points (i.e., 1 second). Adjacent samples do not overlap. The model needs to predict the following 24-step signals based on the previous 76-time steps. We reported the validation mean squared error as the model's prediction performance.

\subsection{Virtual perturbation of ANN models}
For model perturbation, we generated another time series of 100,000 time points, which formed 1,000 samples. During the generation of synthetic data, we added a perturbation with a value of 0.1 to the excitatory interneurons $x_1$ at the $76^{th}$ step of each sample and recorded the changes in the following steps. The perturbation was applied to one region and will affect other regions. The real EC was calculated as the average difference between the following generated data with and without perturbation. For the trained ANNs, we input the perturbed data (time steps 1-76 of each sample) and obtained the predicted signals of the following time steps. The estimated EC was calculated from the average difference between the predicted data with and without perturbation in the input. 


\begin{table}[h]
\centering
\caption{\textbf{Related statistics of 3-channel EEG data prediction.} Five ANN models, including CNN, RNN, GRU, LSTM, and Transformer with 8, 32, 128, and 512 hidden units, were trained to predict EEG signals, respectively. The prediction error and the correlation between the real EC and the NPI-EC were calculated with the test data.}
\begin{tabular}{cccc}
\hline\hline
 \begin{tabular}[c]{@{}c@{}}ANN\\ model\end{tabular}                 & \multicolumn{1}{c}{\begin{tabular}[c]{@{}c@{}}Hidden\\ Units\end{tabular}} & \multicolumn{1}{c}{\begin{tabular}[c]{@{}c@{}}Prediction\\ Error$\downarrow$\end{tabular}} & \multicolumn{1}{c}{\begin{tabular}[c]{@{}c@{}}EC\\ Correlation$\uparrow$\end{tabular}} \\ \hline\hline
\multirow{4}{*}{CNN} & 8   & 5.4477    & 0.7442      \\ \cline{2-4} 
& 32  & 5.4032   & 0.7274                   \\ \cline{2-4} 
& 128  & \textbf{5.3920}  &  \textbf{0.8785}                 \\ \cline{2-4} 
& 512  & 5.4202  &  0.7451   \\ 
\hline\hline
\multirow{4}{*}{RNN} & 8  & \textbf{5.3026}  & -0.1080    
\\ \cline{2-4}  & 32  & 5.3057  & -0.0890  \\ \cline{2-4} 
& 128  & 5.3047 &  \textbf{0.2636}  \\ \cline{2-4}
& 512  & 5.3514   & 0.1820                  \\ \hline\hline
\multirow{4}{*}{LSTM} & 8  & 5.6079   & 0.2593    
\\ \cline{2-4}   & 32  & \textbf{5.4021}   & 0.5353 \\ \cline{2-4} 
& 128  & 5.5618   & 0.2984    \\ \cline{2-4}
& 512  & 5.6119   & \textbf{0.6430}   \\ \hline\hline
\multirow{4}{*}{GRU} & 8  & 5.5250 & 0.3675    
\\ \cline{2-4} & 32  & \textbf{5.3751}   & 0.6186  \\ \cline{2-4} 
& 128  & 5.3816  & \textbf{0.6436}   \\ \cline{2-4}
& 512  & 5.5334  & 0.4198  \\ 
\hline\hline
\multirow{4}{*}{\tiny{Transformer}} & 8  & 5.4221  & 0.7680    
\\ \cline{2-4}  & 32  & 5.4861  & 0.6780   \\ \cline{2-4} 
& 128  & \textbf{5.3803}  & \textbf{0.8110}   \\ \cline{2-4}
& 512  & 5.4132   & 0.8022    \\
\hline\hline
\end{tabular}
\label{table: model performance}
\end{table}

\section{Results}
\subsection{Results on 3-channel synthetic EEG}

We first examined the model performance for EC estimation with 3-channel synthetic EEG data. The model performance was evaluated based on two metrics: time series prediction error and correlation between the real EC and the predicted EC on testing data, as shown in Table~\ref{table: model performance}. The time series prediction error is calculated as the Mean Square Error (MSE) between the real signal and the predicted signal of the last 24 time steps in each sample. EC correlation is defined as the Pearson correlation coefficient between the real EC and the NPI-EC without considering matrix diagonals. We reported the performance of CNN, RNN, LSTM, GRU, and Transformer models with different hidden dimensions. CNN model with 128 hidden dimensions achieved the best EC correlation of 0.8785. The transformer model follows with an EC correlation of 0.8110. For these two models, the hidden dimension with a higher EC correlation also accompanies a lower time series prediction error. LSTM and GRU obtained an inferior EC correlation of 0.6430 and 0.6436, respectively. Vanilla RNN obtained the worst EC correlation of 0.2636. The results indicated that recurrent neural networks are worse than CNN and Transformer in recovering the underlying causal interactions from the synthetic EEG data, although they all achieved comparable time series prediction errors.

We visualized the real EC and the EC inferred from CNN at time step 3 (30 ms) after perturbation (Fig.~\ref{fig: three node result}d, left and middle columns). The inferred EC faithfully recovers the causal interaction from node 0 to nodes 1 and 2. We also visualized the EC estimated by Granger causality (Fig.~\ref{fig: three node result}d, right column). We used multivariate GC to calculate the direct connection between two channels and choose 12 as the input for maxlag based on the minimum value of bic. There are small false positive connections between node 1 and node 2 in EC estimated by GC, which is better suppressed in the NPI-EC. 

\subsection{Results on whole-brain synthetic EEG}

For the whole-brain synthetic EEG with 90 regions, CNN obtained the highest correlation between the real EC and the NPI-EC ($R=0.3340$), compared with Transformer ($R=0.3245$), LSTM ($R=0.2055$), GRU ($R=-0.0096$) and RNN ($R=-0.0006$). This trend is similar to that of 3-channel synthetic data. GRU and RNN wrongly estimate the causality.  GRU's failure may be due to the simpler design of the gating mechanism in contrast to LSTM. 

\begin{figure}[H]
\centering
\includegraphics[width=1\linewidth]{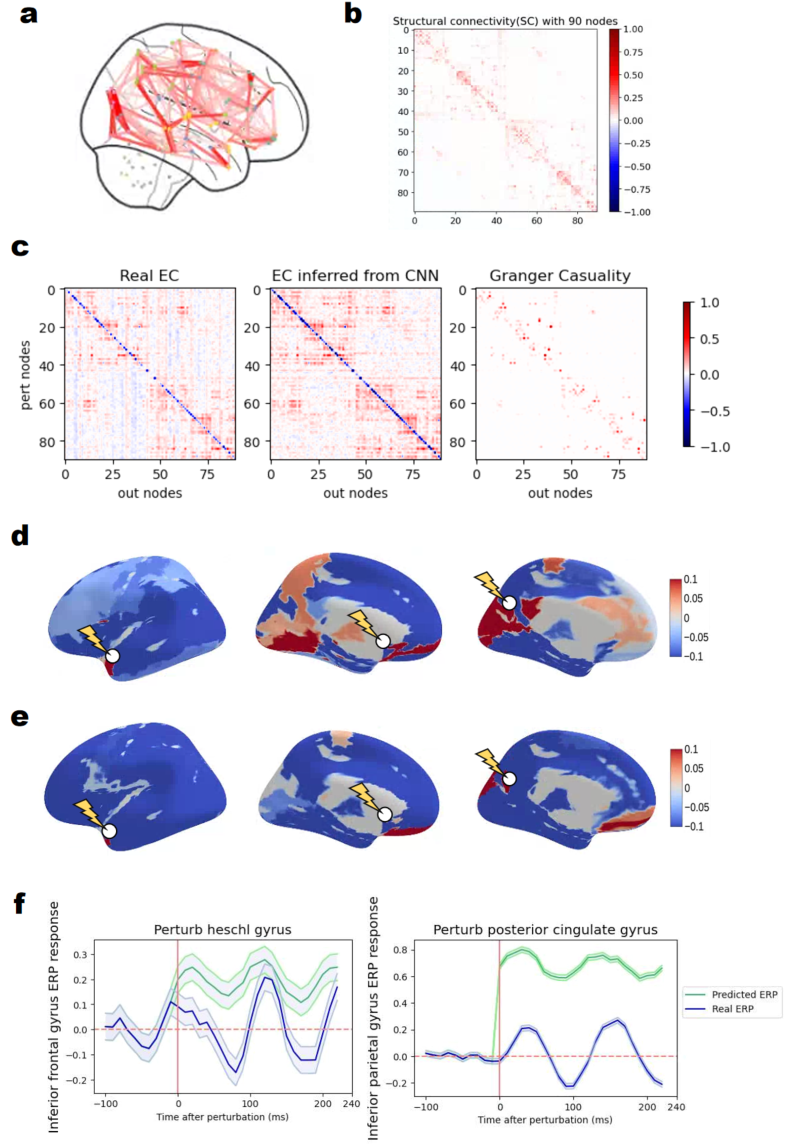}
\caption{\textbf{Data and results visualization of 90-channel synthetic EEG}. \textbf{a}, The real structural connectivity measured from diffusion tensor imaging (mapped on 90 brain regions). 
\textbf{b}, The structural connectivity matrix. \textbf{c}, The comparison among the real EC (left), the EC inferred by perturbing the CNN model (middle), and EC inferred by Granger causality (right). All the EC matrices were re-scaled to the range 0-1 for visualization. \textbf{d}, Spatial distribution of the real EC (i.e., neural responses) by perturbing left amygdala (left), left rectus (middle), and left cuneus (right). \textbf{e}, Spatial distribution of the NPI-EC by perturbing the three regions same as in \textbf{d}. The perturbed region is indicated with an arrow in each panel. \textbf{f}, Sample event-related potential (ERP) of the stimulus-evoked neural responses. ERP is the average response to $1000$ times perturbation. Perturbation is given at time point 0. We show the predicted and real response of the inferior frontal gyrus to heschl gyrus perturbation (left) and the response of the inferior parietal gyrus to the posterior cingulate gyrus perturbation (right).}
\label{fig: 90 node result}
\end{figure}

We visualized the real EC, the EC inferred by CNN and Granger causality at time step 16 (160 ms) after perturbation (Fig.~\ref{fig: 90 node result}c, from left to right). The inferred EC can recover the real EC faithfully, with a high EC correlation ($R=0.7081$) (for this time step). In contrast, the Granger causality inferred EC is much worse ($R=0.3136$).

To exhibit the effect of perturbing one region on the others, we show the spatial distribution of signal changes resulting from perturbing a specific seed region in the J-R model (i.e., real EC) in Fig.~\ref{fig: 90 node result}d, as well as the NPI-EC of CNN model in Fig.~\ref{fig: 90 node result}e. The NPI-EC recovers the general distribution of real EC. To show the temporal evolution of the signals after perturbation, we also compare the real event-related potentials (ERP) and the predicted ERP under perturbation in Fig.~\ref{fig: 90 node result}f. After the perturbation was given at time point 0, the predicted ERP and the real ERP show a similar trend of change, although their exact values were different.


\section{Discussion}

In this study, we presented a testbed for perturbation-based EC estimation methods with synthetic EEG data. Our results validated that by perturbing specific types of ANN prediction models (i.e., CNN and Transformer), we can estimate the underlying causal interactions among different nodes effectively. 

ANN models have been widely used in time series forecasting and achieved SOTA performance. However, it is unclear whether the models can reveal the causal interactions among different variates. Our experiments showed that specific types of models can encapsulate the underlying causal interactions of synthetic EEG data. CNN and Transformer achieved higher performance here, probably due to they can capture the oscillation characteristics in synthetic EEG data~\cite{lawhern2018eegnet,song2022eeg}.

In future studies, several important questions remain to be investigated. Firstly, what are the effects of different types of perturbation? Some perturbations may cause the signals to be outside the manifold of natural signals, while others may not~\cite{shenoy2021measurement}. Specific forms of perturbations may resemble those in real brain stimulation. It is critical to investigate these different types of perturbations on EC estimation. Secondly, it still lacks a clear explanation of why CNN and Transformer work better on EC estimation than RNN-series models. Do they also work well on real EEG data? How to choose the proper model for different types of data? These questions need to be further investigated in the future.




\section*{Acknowledgements}
We thank Mr. Zhichao Liang for sharing some code, and Mr. Song Wang, and Mr. Kaining Peng for their useful discussions. This work was funded in part by Shenzhen Science and Technology Innovation Committee (2022410129, 20200925155957004, KCXFZ2020122117340001, JCYJ20220818100213029, SGDX2020110309280100), Guangdong Provincial Key Laboratory of Advanced Biomaterials (2022B1212010003).
0
\section*{Appendix}

\textbf{Convolutional Neural Network (CNN).} The Convolutional Neural Network is a feedforward multilayered hierarchical network, which is a widely used ANN model. It uses a combination of convolutional layers, nonlinear processing units, and subsampling layers to automatically extract features from the raw pixel data of the image for improved categorization with 2-dimensional data. It can be applied to 1-dimensional time series data by temporal convolution.

\textbf{Vanilla RNN.} Vanilla RNN was used as an example of a simple nonlinear forecasting model. The iteration of the hidden state is represented as
\begin{equation}
h_t=tanh(x_tW_{ih}^T+b_{ih}+h_{t-1}W_{hh}^T+b_{hh}), 
\end{equation}
where $x_t$ is the input and $h_t$ is the hidden state.

\textbf{Long Short-Term Memory (LSTM).} LSTM adds gate design to vanilla RNN to capture long-term dependencies in the time series. The detailed computation in an LSTM unit is shown below:

\begin{equation}
\begin{aligned}
i_t & =\sigma\left(W_{i i} x_t+b_{i i}+W_{h i} h_{t-1}+b_{h i}\right) \\
f_t & =\sigma\left(W_{i f} x_t+b_{i f}+W_{h f} h_{t-1}+b_{h f}\right) \\
g_t & =\tanh \left(W_{i g} x_t+b_{i g}+W_{h g} h_{t-1}+b_{h g}\right) \\
o_t & =\sigma\left(W_{i o} x_t+b_{i o}+W_{h o} h_{t-1}+b_{h o}\right) \\
c_t & =f_t \odot c_{t-1}+i_t \odot g_t \\
h_t & =o_t \odot \tanh \left(c_t\right)
\end{aligned}
\end{equation}
where $x_t$ is the input and $h_t$ is the hidden state. $i_t$, $f_t$, and $o_t$ represent the output of the input gate, the forget gate, and the output gate, respectively. $g_t$ is the candidate cell state and $c_t$ is the cell state.

\textbf{Gated Recurrent Unit (GRU).} GRU simplified the gate design in LSTM to improve the computational efficiency:

\begin{equation}
\begin{aligned}
& r_t=\sigma\left(W_{i r} x_t+b_{i r}+W_{h r} h_{(t-1)}+b_{h r}\right) \\
& z_t=\sigma\left(W_{i z} x_t+b_{i z}+W_{h z} h_{(t-1)}+b_{h z}\right) \\
& n_t=\tanh \left(W_{i n} x_t+b_{i n}+r_t *\left(W_{h n} h_{(t-1)}+b_{h n}\right)\right) \\
& h_t=\left(1-z_t\right) * n_t+z_t * h_{(t-1)}
\end{aligned}
\end{equation}
where $x_t$ is the input and $h_t$ is the hidden state. $r_t$ and $z_t$ represent the output of the reset gate and the update gate, respectively. $n_t$ is the candidate's hidden state.

\textbf{Transformer.}
The Transformer model employs a self-attention mechanism. By encoding the input time series into a set of vectors and applying self-attention across all time steps, the Transformer model captures both local and global dependencies, enabling accurate predictions. The attention weights are obtained through a softmax function applied to the scaled dot-product of query, key, and value embeddings. It computes the attention function on a set of queries simultaneously, packed together into a matrix Q. The keys and values are also packed together into matrices K and V. The output is computed as 
\begin{equation}
\operatorname{Attention}(Q, K, V)=\operatorname{softmax}\left(\frac{Q K^T}{\sqrt{d_k}}\right) V ,
\end{equation}
where $d_k$ is the dimension of queries and keys. By iteratively updating the hidden states through the self-attention layers, the Transformer model learns to capture complex temporal dependencies, facilitating accurate prediction of future values in the time series.

where $\mathbf{y}_t$ is a vector of observed variables at time $t$, $\mathbf{c}$ is a constant vector, $\mathbf{A}i$ is the coefficient matrix associated with the $i$-th lag, $\mathbf{y}{t-i}$ represents the vector of lagged variables, and $\mathbf{e}_t$ is a vector of error terms assumed to be white noise. 

\bibliographystyle{named}
\bibliography{ref}

\end{document}